\documentclass[a4paper,fleqn]{cas-sc}
\usepackage[numbers,sort&compress]{natbib} 
\pdfoutput=1 %PDFLaTeX processing
 
\usepackage{overpic} 

\usepackage{threeparttable}

\def\tsc#1{\csdef{#1}{\textsc{\lowercase{#1}}\xspace}}
\tsc{WGM}
\tsc{QE}
\tsc{EP}
\tsc{PMS}
\tsc{BEC}
\tsc{DE}
\begin{document}
\let\WriteBookmarks\relax
\def\floatpagepagefraction{1}
\def\textpagefraction{.001}

\shorttitle{A fluid flow model for the pressure loss through perforated plates}

\shortauthors{S Li et~al.}

\title [mode = title]{A fluid flow model for the pressure loss through perforated plates}

\author[1]{Shuai Li}

\cormark[1]

\ead{shuai.li@chalmers.se}

\credit{Conceptualization, Methodology, Software, Validation, Formal analysis, Investigation, Data curation, Writing - original draft, Writing - review \& editing, Visualization}

\address[1]{Department of Mechanics and Maritime Sciences, Chalmers University of Technology, SE-412 96 Gothenburg, Sweden}

\author[1]{Lars Davidson}[]
 
\ead{lars.davidson@chalmers.se}

\credit{Formal analysis, Writing - review \& editing, Visualization, Supervision, Funding acquisition}

\author%
[1,2]
{Shia-Hui Peng}

\ead{peng@chalmers.se}

\credit{Formal analysis, Writing - review \& editing, Visualization, Supervision, Funding acquisition}
\address[2]{FOI - Swedish Defense Research Agency, SE-164 90 Stockholm, Sweden}

\cortext[cor1]{Corresponding author}

\begin{abstract}
A fluid flow through a perforated plate is a common problem in a wide variety of practical applications in thermal, mechanical, chemical, civil, nuclear, ocean and aerospace engineering.  In this paper, we proposed a novel fluid flow model for the pressure loss through plates with circular perforations in both laminar and turbulent flows. The design of this model is based on the recent measurements conducted at ONERA in the framework of the on-going European Union H2020 INVENTOR project, as well as an existing model for laminar flows. The new model is then validated against existing numerical simulations in the laminar regime and experiments in the turbulent regime. Overall, the predictions given by the new model agree well with the numerical simulations and experiments, and are superior to other models in the literature. This is significant, considering that the present model is much simpler than these previous models. To demonstrate the application of the new model in numerical simulations, two-dimensional channel flows are simulated using Reynolds-averaged Navier–Stokes (RANS) equations with the new model as a pressure-drop source term added to the momentum equations. Results show that the RANS predictions agree very well with the present model predictions.
\end{abstract}

% Keywords

% Disable the thumbnails
\ExplSyntaxOn
\keys_set:nn { stm / mktitle } { nologo }
\ExplSyntaxOff

\maketitle

\section{Introduction}

A fluid flow through a perforated plate is a common problem in a wide variety of practical applications in thermal, mechanical, chemical, civil, nuclear, ocean and aerospace engineering. To enumerate a few, perforated plates are used for the design of heat transfer devices \cite{ahmimache2022heat, lee2002heat, mcmahon1950perforated, shevyakova1983study, kutscher1994heat, white2010experimentally, tomic2014methodology, arghode2015experimental, raju2017heat, tomic2018perforated, husin2021modification} and flow conditioning devices \cite{laws1995further, spearman1996comparison, laribi2003comparative, xiong2003velocity, hoffmann2011effect, yaici20143d, laribi2015discharge}, the control of flames in combustion chambers \cite{noiray2007passive, oh2016stabilization, rashwan2017experimental, kim2020effects, younesian2021experimental, younesian2022visualization} and turbulence \cite{naot1980penetration, liu2004generation, liu2007turbulent, dhineshkumar2013large}, the reduction of aerodynamic noise \cite{sakaliyski2007aero, rubio2019mechanisms, laffay2020experimental, sumesh2021aerodynamic}, and so forth. The flow through a perforated plate usually features flow separation near the leading edge of the pores, contraction until the vena contracta is reached, and expansion after the vena contracta. The major pressure losses are induced by the flow expansion process during which considerable energy dissipation takes place \cite{miller1990internal}.

In heat transfer design, flow maldistribution is a major issue degrading the performance of heat transfer devices \cite{yaici20143d, pacio2010study, nielsen2013influence, beckedorff2022flow}. It is common to make use of perforated plates to provide a more uniform flow velocity. As such, it is necessary to accurately predict the pressure drop across the perforated plate to ensure that it is sufficient to provide good flow uniformity. Moreover, perforated plates can also enhance the turbulence homogeneity so as to improve the system performance of heat transfer devices. However, perforated plates also cause drag increases, leading to more energy consumption of the system. It is thus of great significance to study the pressure losses across perforated plates, and to develop empirical models to estimate a priori the pressure losses for a better compromise between flow conditioning and pressure loss. 

\begin{figure}
\centering
\begin{overpic}[width=0.3\textwidth,trim=4.5cm -2cm 0cm 0cmm, clip]{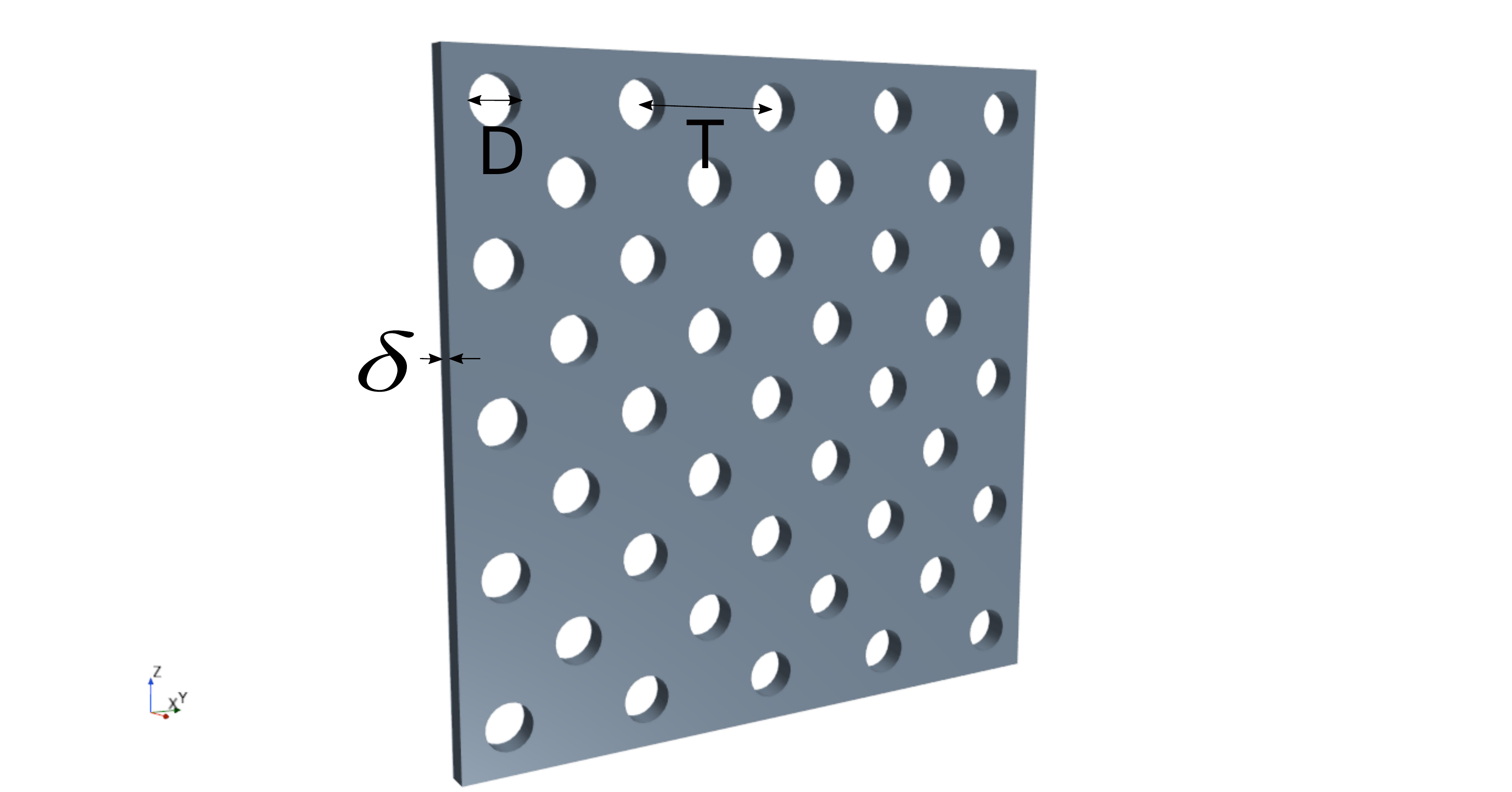} \put(5,68){($a$)} \end{overpic} 
\begin{overpic}[width=0.45\textwidth,trim=0cm 3.5cm 0cm -2cm, clip]{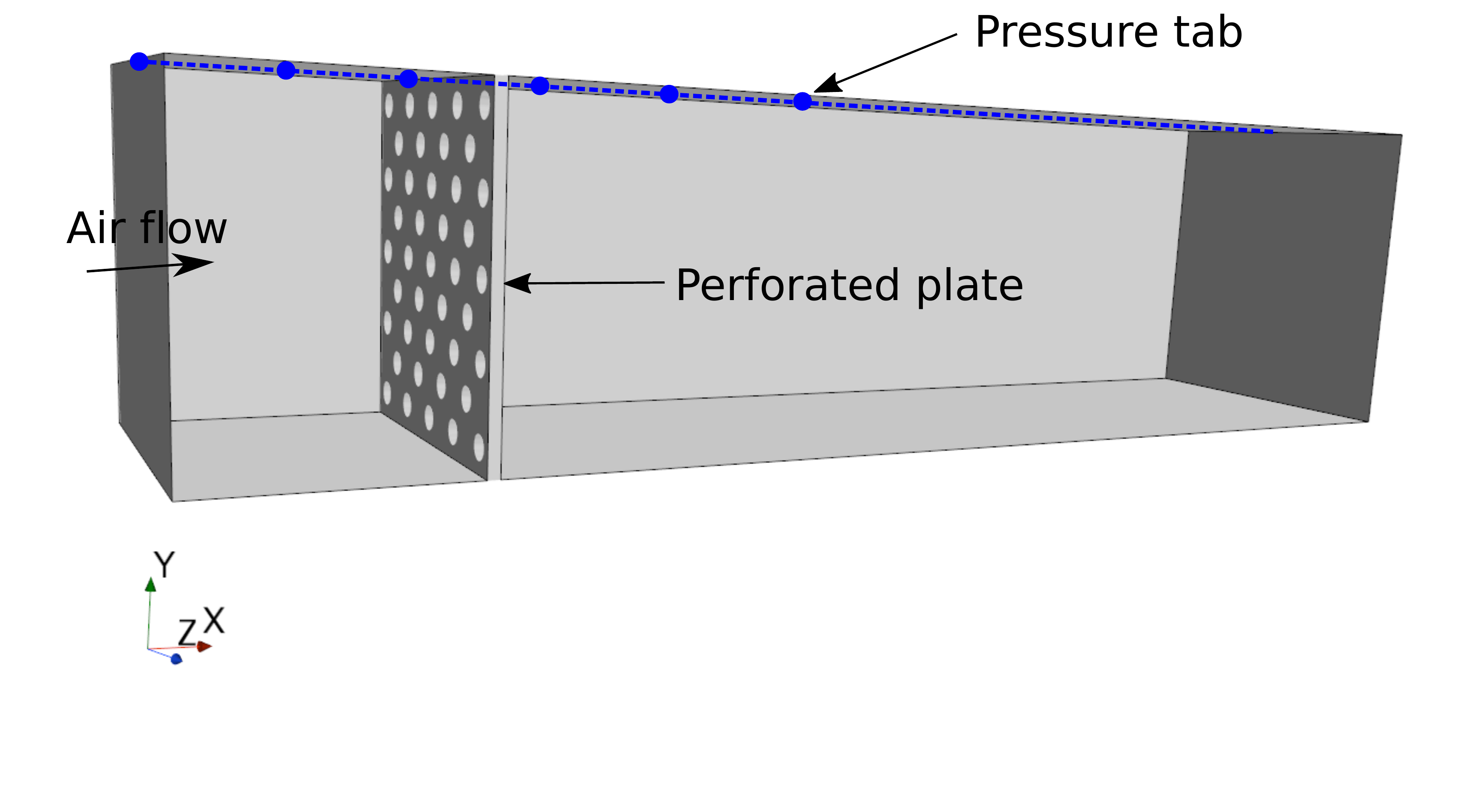} \put(0,45){($b$)} \end{overpic} 
\\
\caption{Sketches of (a) a perforated plate; and (b) the B2A wind-tunnel test section.}
\label{fig: sketch_perforated plate_B2A}
\end{figure}

Previous research on pressure losses through perforated plates can primarily be categorized into experimental and numerical studies. Yavuzkurt \& Catchen \cite{yavuzkurt2003dependence} studied experimentally the dependence of pressure loss on air speed with a set of wind-tunnel tests on seven perforated plates of different thicknesses and porosities at different air speeds up to $76$ m/s. It was found that, for most of the plates, the pressure loss increases linearly with the square of the air speed, indicating the dominance of inertial pressure losses. For a specific type of plates, small deviations from this linear behavior can occur, due to a type of wake characterized by large-scale flow oscillations downstream of the plate. Recently, Méry \& Sebbane \cite{mery2023aerodynamic} investigated experimentally the pressure loss characteristics of plates with circular perforations of different pore diameter, spacing and plate thickness in the framework of the on-going European Union H2020 INVENTOR project. Fig. \ref{fig: sketch_perforated plate_B2A}(a) shows a sketch of a perforated plate with a thickness of $\delta$. The pores have a diameter of $D$ and  are separated by a spacing of $T$. Different pore diameter and spacing of perforated plates lead to different porosity $\varepsilon$ which is defined as the ratio between the area of pores and the total area of the perforated plate. It has been shown that the pressure drop coefficient (i.e. pressure drop normalized by the dynamic pressure) is independent of air speed but varies significantly with the plate thickness ($\delta/D$). Through numerical simulations, Bayazit et al. \cite{bayazit2014perforated} investigated the pressure losses across perforated plates of different thicknesses ($\delta/D=0.5$ and $1.0$) and porosities ($\varepsilon=0.2$, $0.35$ and $0.5$) at a wide variety of Reynolds numbers ranging from laminar to turbulent regimes. It was shown that, for laminar flows, higher pressure drops were associated with thicker plates and lower porosities. For turbulent flows, however, the thinner plate caused higher pressure drops as did lower porosities. These trend reversals result from the differences in separated-flow reattachment patterns for the different plate thicknesses in different flow regimes. Bae \& Kim \cite{bae2016numerical} numerically studied the pressure losses of laminar flows through thick perforated plates ($\delta/D \geq 1$) at pore-level Reynolds numbers up to 25 and proposed a model, in a mathematically simple form, to predict the pressure loss in laminar regimes. This model was well validated against the experiment but is only applicable to laminar flows. Similarly, based on experiments, a number of models for the pressure loss were proposed by, for example, Idelichik \cite{idelchik1986handbook}, Idelichik \cite{idelchik1994handbook}, Miller \cite{miller1990internal}, Kast \cite{kast2010pressure}, Holt et al. \cite{holt2011cavitation}, ESDU \cite{engineering1981flow} and Malavasi et al. \cite{malavasi2012pressure}, etc. However, these models are much more complicated including parameters to determine using empirical formulae, and are only applicable to high-Reynolds-number turbulent flows. The purpose of the present study is to design a novel model, in a mathematically simple form, but with high predictive accuracies, to characterize the pressure loss through perforated plates in both laminar and turbulent regimes.

In this paper, we construct a correction factor for the Forchheimer coefficient of the Bae \& Kim model  \cite{bae2016numerical}, based on the experiment of Méry \& Sebbane \cite{mery2023aerodynamic}, formulating a new model valid for both laminar and turbulent flow regimes. The new model is then validated against existing simulation and experimental data in the literature, with comparisons with some other popular models. Furthermore, to showcase an application of the new model in numerical simulations, we perform two-dimensional channel-flow simulations using RANS with the new model as a pressure-drop source term added to the momentum equations. The RANS-predicted pressure losses are validated against the model-predicted pressure losses, demonstrating good prediction capabilities of the new model.

The paper is organized as follows: \S\ref{experimental_data} is devoted to a description of the experimental data deployed for the development and validation of the present model for plates with circular perforations. In \S\ref{perforated_plate_models}, a discussion of existing perforated-plate models is presented and then a novel model is proposed. In \S\ref{results}, the proposed perforated-plate model is validated against existing numerical simulations in the laminar regime and experiments in the turbulent regime. Then, \S\ref{application_model} showcases application of the proposed model to numerical simulations. Finally, concluding remarks are summarized in \S\ref{conclusion}.

\section{The experimental data}\label{experimental_data}

In the present study, the experimental data of Méry \& Sebbane \cite{mery2023aerodynamic} and Yavuzkurt \& Catchen \cite{yavuzkurt2003dependence} are used to develop and validate, respectively, the novel model for the pressure loss through perforated plates. The measurements were conducted at the Aero-thermo-Acoustics Bench (B2A) of the Office National d'Etudes et de Recherches Aérospatiales (ONERA) and at the Pennsylvania State University, respectively. 

The B2A wind tunnel is designed so that the static flow temperature can be accurately regulated from the ambient temperature up to $300^\circ C$, with a mean flow bulk Mach number up to 0.5. A 0.2-m-long test section is equipped with two silica windows for optical access. This wind-tunnel cross section has a dimension of 50 mm $\times$ 50 mm. A sketch of the wind-tunnel test section is shown in Fig. \ref{fig: sketch_perforated plate_B2A}(b). More details on the B2A wind tunnel can be found in Ref.~\cite{minotti2008characterization, mery2023aerodynamic}. It is important to note that, in this experiment, the perforated plate fully covers the entire cross section. In other words, the perforated plate has the same dimension as the wind-tunnel cross section, i.e. 50 mm in width and span. The pressure drop was measured by static pressure taps on the top of the test section upstream and downstream the perforated plate. In the experiments, the airflow speeds at the wind-tunnel entrance are $U_{0} = 16.6$,  $25$ and $35$ m/s, respectively. These measurements were recently conducted in the framework of the on-going European Union H2020 INVENTOR project.

The measurements of Yavuzkurt \& Catchen \cite{yavuzkurt2003dependence} were conducted in a pipe-flow apparatus instead of a channel-flow wind tunnel. The internal diameter of the pipe is $90.17$ mm. The airflow was provided by a 5 hp blower operated in the suction mode and was controlled using a bypass and an inline valve respectively. The pressure taps were placed $710$ mm upstream and downstream from the plate to reduce the effects of local turbulence occurring near the perforated plate. The anemometer used to measure the airflow speed was placed near the upstream opening of the entrance pipe, which provided a relatively uniform distribution of flow velocities across the inlet area of the pipe. The pressure losses were measured over a range from $0$ Pa to $3500$ Pa. More details on the experimental apparatus can be found in Ref.~\cite{yavuzkurt2003dependence}.

Table \ref{tab:Mery_Sebbane} shows the pore size $D$, pore spacing $T$, plate thickness $\delta$ and porosity $\varepsilon$ of the four perforated plates deployed in the B2A wind-tunnel experiments \cite{mery2023aerodynamic}. The porosity, defined as the ratio between the aera of pores and the total area of the perforated plate, can be calculated using the pore size and the pore spacing. Similarly, Table \ref{tab:Yuvuzkurt_Catchen} shows the parameters of the seven perforated plates in the experiments of Yavuzkurt \& Catchen \cite{yavuzkurt2003dependence}. The pressure losses measured in Méry \& Sebbane \cite{mery2023aerodynamic} and Yavuzkurt \& Catchen \cite{yavuzkurt2003dependence} will be plotted in the following Section \ref{perforated_plate_models} and Section \ref{results}.

\begin{table}[width=.47\linewidth,cols=4,pos=h]
\begin{threeparttable}
\caption{Parameters of the perforated plates used in the experiment of Méry \& Sebbane \cite{mery2023aerodynamic}.}\label{tab:Mery_Sebbane}
\begin{tabular*}{\tblwidth}{@{} cccccc@{} }
\toprule
      Plate  & $D$ (mm) & $T$ (mm) & $\delta$ (mm) & $\varepsilon$ & $\delta/D$ \\[3pt]
\midrule
       1   & 5.0 & 6.0 & 1.0 & 0.623 & 0.20 \\
       2   & 4.0 & 6.0 & 1.0 & 0.403 & 0.25 \\       
       3   & 2.0 & 3.0 & 2.0 & 0.403 & 1.00 \\
       4   & 2.0 & 3.0 & 1.0 & 0.403 & 0.50 \\
\bottomrule
\end{tabular*}
\end{threeparttable}
\end{table}

\begin{table}[width=.48\linewidth,cols=4,pos=h]
\begin{threeparttable}
\caption{Parameters of the perforated plates used in the experiment of Yavuzkurt \& Catchen \cite{yavuzkurt2003dependence}.}\label{tab:Yuvuzkurt_Catchen}
\begin{tabular*}{\tblwidth}{@{} cccccc@{} }
\toprule
      Plate  & $D$ (mm) & $T$ (mm) & $\delta$ (mm) & $\varepsilon$ & $\delta/D$ \\[3pt]
\midrule
       1   & 1.60 & 3.18 & 1.02 & 0.227 & 0.638 \\
       2   & 1.91 & 2.54 & 0.81 & 0.510 & 0.424 \\       
       3   & 2.38 & 6.88 & 0.91 & 0.109 & 0.382 \\
       4   & 2.78 & 4.76 & 0.81 & 0.309 & 0.291 \\
       5   & 3.18 & 4.76 & 0.91 & 0.403 & 0.286 \\       
       6   & 4.76 & 9.53 & 0.97 & 0.227 & 0.204 \\
       7   & 3.18 & 4.76 & 3.18 & 0.403 & 1.000 \\
\bottomrule
\end{tabular*}
\end{threeparttable}
\end{table}

\section{Empirical models for the pressure loss through perforated plates}\label{perforated_plate_models}

The pressure losses caused by flow-through perforated plates are often characterized by the Darcy-Forchheimer equation, as follows
\begin {equation}\label{pressure_gradient}
-\nabla P = \frac{\mu}{K}U + \rho\alpha\lvert U \rvert U
\end {equation}
where $K$ is the permeability of the porous medium, $\alpha$ is the Forchheimer coefficient or non-Darcy coefficient, $\rho$ is the fluid density, and $\mu$ is the dynamic viscosity of the fluid. The Darcy term, i.e. the first term on the right-hand side (RHS) of Eq. (\ref{pressure_gradient}), corresponds to the viscous pressure loss while the Forchheimer term, i.e. the second term on the RHS of Eq. (\ref{pressure_gradient}), represents the inertial pressure loss. For cases where the pore-level Reynolds number is sufficiently low, the Darcy term strongly dominates over the Forchheimer term \cite{bae2016numerical, tanner2019flow, shahzad2022permeability, li2023numerical}. 

The pressure gradient ($-\nabla P$) on the left-hand side (LHS) of Eq. (\ref{pressure_gradient}) can also be expressed using the pressure drop ($\Delta p$) across the plate with a thickness of $\delta$, as follows
\begin {equation}\label{pressure_gradient_to_drop}
-\nabla P = \frac{\Delta p}{\delta}.
\end {equation}
Then, the Darcy-Forchheimer drag obtained from Eq. (\ref{pressure_gradient}) can be written as \cite{lee1997modeling}
\begin {equation}\label{pressure_drop}
\frac{\Delta p D^2}{\mu \delta U} = \frac{D^2}{K} + \varepsilon \alpha D Re_p,
\end {equation}
where $Re_p=\rho UD/(\mu \varepsilon)$ is the pore-level Reynolds number. From Eq. (\ref{pressure_drop}), the normalized pressure drop through perforated plates can be expressed as
\begin {equation}\label{pressure_coef}
\frac{\Delta p}{\rho U^2} = \frac{\delta D}{K \varepsilon Re_p} + \alpha \delta.
\end {equation}
The first term on the RHS of Eq. (\ref{pressure_coef}) is the contribution of the Darcy term to the normalized pressure drop and depends on both the plate characteristics (permeability, porosities, thickness, etc.) and the flow characteristics (fluid viscosity, flow speed or Reynolds number). The second term in the RHS of Eq. (\ref{pressure_coef}) is the Forchheimer contribution and only depends on the plate characteristics (porosities, thickness, etc.).

It is also interesting to note that, for general porous medium, Vafai \& Tien \cite{vafai1981boundary} included an additional diffusion term ($-\mu_e\nabla^2U$, where $\mu_e$ is the effective viscosity of porous medium) on the RHS of Eq. (\ref{pressure_gradient}) and thus proposed a much more complicated formulation for the pressure loss through perforated plates, i.e. the so-called Brinkman-Forchheimer equation \cite{vafai1981boundary}. This equation is considered a more general description of the flow resistance in porous medium because of the diffusion term (or Brinkman term) accounting for the boundary effect. However, this additional diffusion term is less significant for perforated plates \cite{bae2016numerical} and is neglected in this study, as in many other previous studies \cite{bae2016numerical, tanner2019flow, shahzad2022permeability}.

\subsection{The Bae \& Kim model} \label{Bae_Kim_model}

Although Eq. (\ref{pressure_coef}) provides a mathematical description of the pressure loss through perforated plates, the permeability $K$ and Forchheimer coefficient $\alpha$ are yet to be determined. On the basis of the numerical simulations of laminar flows ($Re_p<25$), Bae \& Kim \cite{bae2016numerical} developed the following model to estimate $K$ and $\alpha$ of the perforated plate.
\begin {equation}\label{parameters_BaeKim2016}
K = \frac{\varepsilon D^2 \delta}{32\delta+15D}, \:\:\: \alpha = \frac{3(1-\varepsilon)}{4\varepsilon^2\delta}
\end {equation}
This model has been validated against the experiment of Doerffer \& Bohning \cite{doerffer2000modelling}. The reader may refer to Ref. \cite{bae2016numerical} for the detailed comparison between the model predictions and the measurements.

\renewcommand{\figurename}{Fig.}
\begin{figure}
\centerline{\includegraphics[angle=0,width=12.0cm]
  {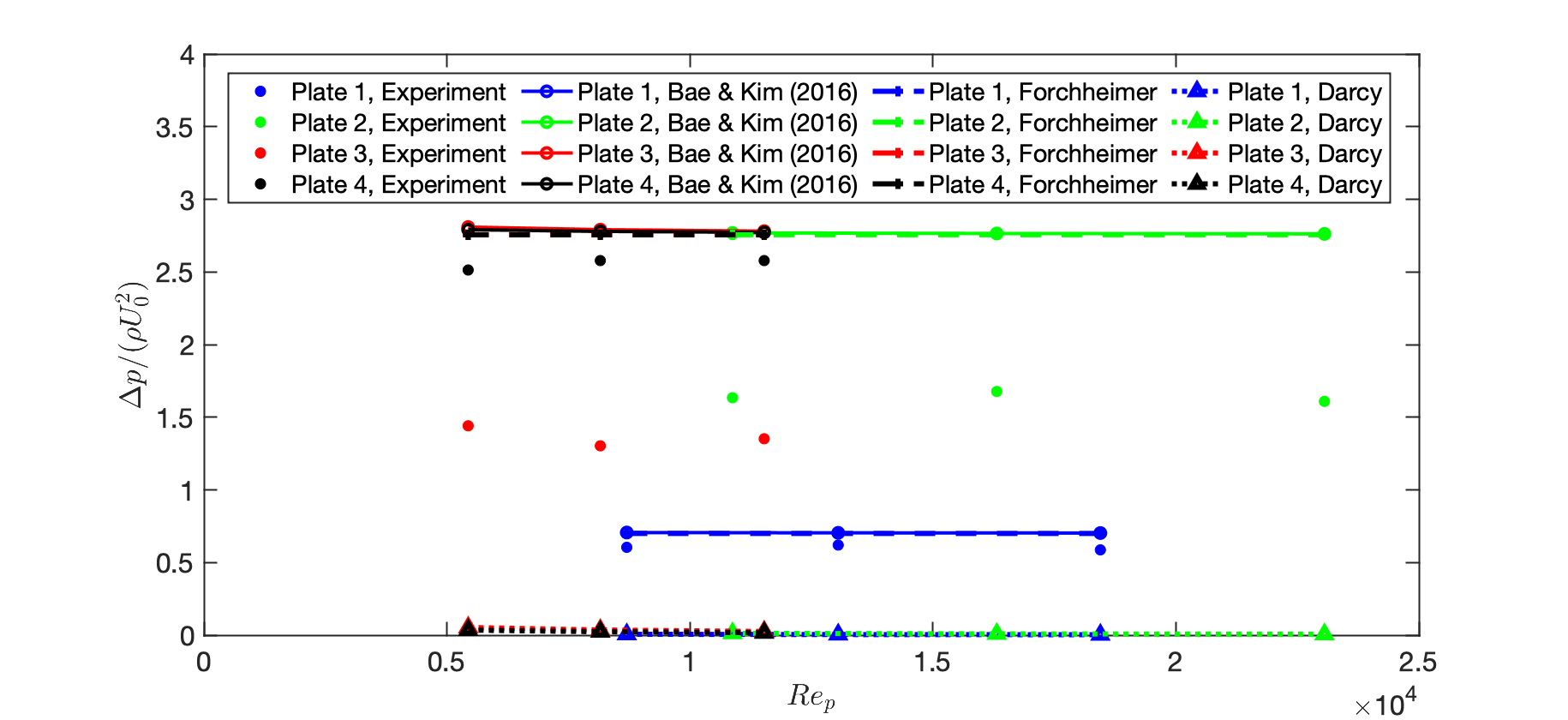}}
\caption{Comparison of the pressure drop predicted by the Bae \& Kim model \cite{bae2016numerical} with the experiment of Méry \& Sebbane \cite{mery2023aerodynamic}. Both Darcy and Forchheimer contribution of the predicted pressure drop are also shown.} 
\label{fig:dp_by_Bae_Kim_16_with_MS23}
\end{figure}

To further assess the accuracy of the Bae \& Kim model \cite{bae2016numerical} in predicting turbulent flows through perforated plates, we apply the model to predict the recent experiment of Méry \& Sebbane \cite{mery2023aerodynamic}. Figure \ref{fig:dp_by_Bae_Kim_16_with_MS23} shows a comparison of pressure drop between the prediction of the Bae \& Kim model \cite{bae2016numerical} and the experiment of Méry \& Sebbane \cite{mery2023aerodynamic}. Although the predicted pressure drops of Plate 1 and 4 are somewhat close to the experiment, those of Plate 2 and 3 are not. It is also clear that the Darcy contribution, i.e. $\delta D/(K \varepsilon Re_p)$, is negligibly small and thus the Forchheimer contribution ($\alpha \delta$) dominates in the present turbulent flow regime. However, it is interesting to note that, at low Reynolds numbers ($Re_p < 25$), conversely, the Darcy contribution dominates over the Forchheimer contribution, as shown in Ref.~\cite{bae2016numerical}. In Fig. 2, the predicted pressure drops of Plate 2, 3 and 4 (whose porosities are the same) are essentially similar due to the fact that the dominant Forchheimer contribution depends only on the plate porosity and not on the plate thickness ratio $\delta/D$. Moreover, it is also found that both experiments and the Bae \& Kim model \cite{bae2016numerical} show that the pressure drops are almost independent of the pore-level Reynolds number $Re_p$, again suggesting that the viscous Darcy contribution is negligibly small and the inertial Forchheimer contribution is dominant in the present flow regime. Overall, although the Bae \& Kim model has been previously shown to perform very well in the prediction of laminar flows through perforated plates \cite{bae2016numerical}, its performance in turbulent flow predictions is not satisfying. Thus, an extension of the Bae \& Kim  model \cite{bae2016numerical} to turbulent regimes is needed.

\renewcommand{\figurename}{Fig.}
\begin{figure}
\centerline{\includegraphics[angle=0,width=18.0cm]
  {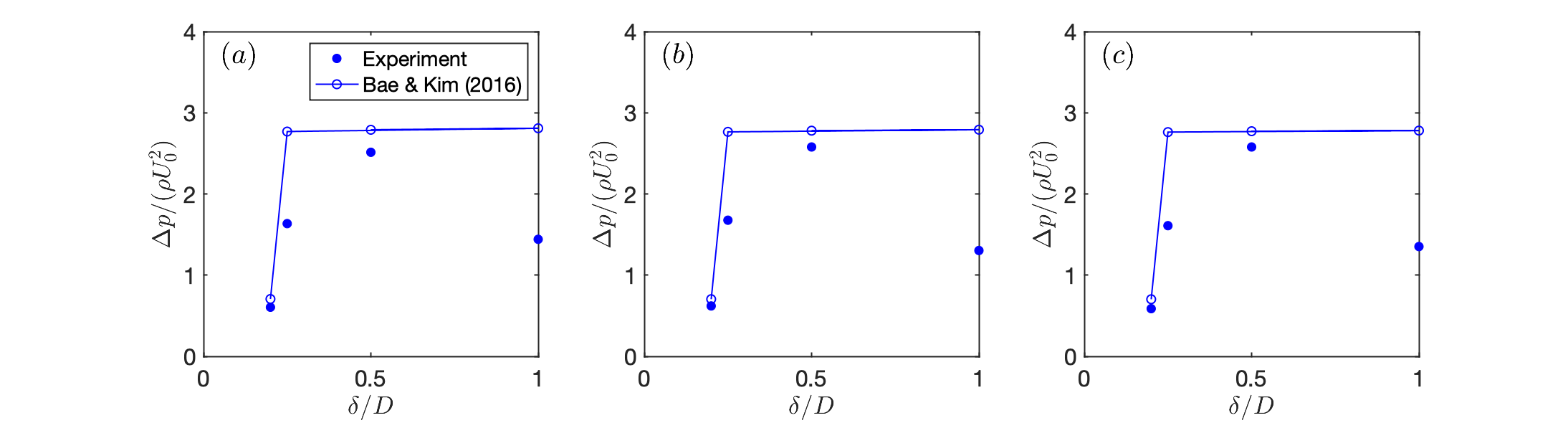}}
\caption{Comparison of the pressure drop predicted by the Bae \& Kim model \cite{bae2016numerical} with the experiment of Méry \& Sebbane \cite{mery2023aerodynamic} at three different flow velocities: (\textit{a}) 16.6 m/s; (\textit{b}) 25 m/s; and (\textit{c}) 35 m/s.} 
\label{fig:dp_by_Bae_Kim_16_at_3_Rep}
\end{figure}

Figure \ref{fig:dp_by_Bae_Kim_16_at_3_Rep} shows a comparison of pressure drop between the prediction of the Bae \& Kim model \cite{bae2016numerical} and the experiment of Méry \& Sebbane \cite{mery2023aerodynamic} at three different flow velocities. For Plate 1 ($\delta/D=0.2$), the predictions agree well with the experiments. For Plates 2, 3 and 4 whose porosities are the same ($\varepsilon = 0.4031$), the measured pressure drops vary with the plate thickness $\delta/D$ whereas the predicted ones do not. This is again because the dominant Forchheimer contribution of the Bae \& Kim  model \cite{bae2016numerical} does not depend on the plate thickness ratio $\delta/D$. Therefore, to extend the Bae \& Kim  model \cite{bae2016numerical} to turbulent regimes, it is reasonable to propose a correction factor that takes into account the effect of plate thickness ratio.

\subsection{An extension of the Bae \& Kim model to turbulent regimes} \label{extension}

Considering the differences between the experimental data and the prediction of the Bae \& Kim model \cite{bae2016numerical}, a correction factor $f$, as a function of the plate porosity $\varepsilon$ and thickness ratio $\delta/D$, is proposed to correct the Forchheimer coefficient $\alpha$ so that the Bae \& Kim model \cite{bae2016numerical} becomes applicable for turbulent regimes while at the same time it remains effective in laminar regimes. Through fitting of the experimental data, we propose a correction factor $f$, as follows

\begin {equation}\label{factor}
f\left(\varepsilon, \frac{\delta}{D}\right) = \frac{0.3}{1-\varepsilon}\left[6\left(\frac{\delta}{D}\right)-5\left(\frac{\delta}{D}\right)^2\right]
\end {equation}

With the factor (Eq. (\ref{factor})) to correct the Forchheimer coefficient, in the new model, the permeability and the Forchheimer coefficient become
\begin {equation}\label{parameters_improved}
K = \frac{\varepsilon D^2 \delta}{32\delta+15D}, \:\:\: \alpha = \frac{9}{40\varepsilon^2\delta}\left[6\left(\frac{\delta}{D}\right)-5\left(\frac{\delta}{D}\right)^2\right]
\end {equation}

The validation of this proposed model is presented in Section \ref{laminar_flow} for laminar flows and in Section \ref{turbulent_flow} for turbulent flows.

\section{Results}\label{results}

In this section, we validate the proposed model (Eq. (\ref{parameters_improved})) against existing numerical simulations in laminar regimes and experiments in turbulent regimes. We also compare our new model with other previous models \cite{idelchik1994handbook, kast2010pressure, miller1990internal, holt2011cavitation} (see Appendix).

\subsection{Laminar flow}\label{laminar_flow}

\renewcommand{\figurename}{Fig.}
\begin{figure}
\centerline{\includegraphics[angle=0,width=12.0cm]
  {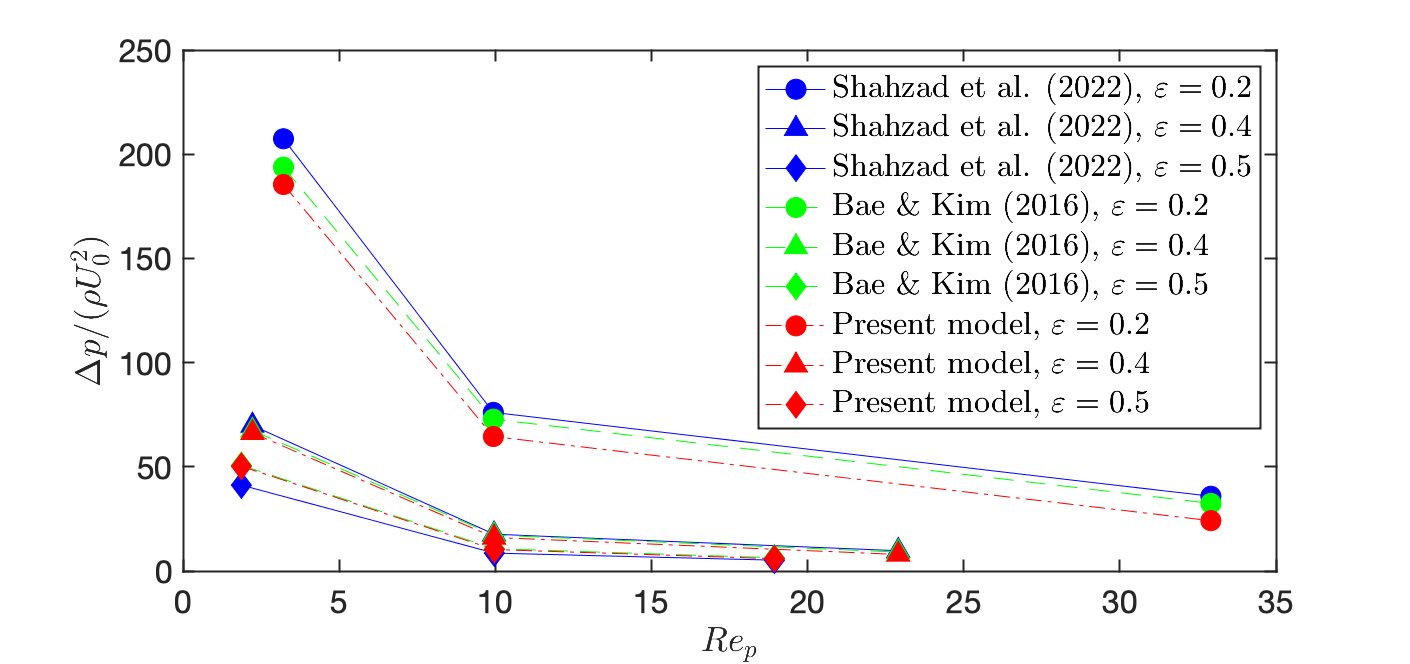}}
\caption{Comparison of the pressure drops predicted by the present model (Eq. (\ref{parameters_improved})) and the Bae \& Kim model \cite{bae2016numerical} with the numerical simulations of Shahzad et al. \cite{shahzad2022permeability}.} 
\label{fig:laminar_dp_with_Shahzad22}
\end{figure}

Shahzad et al. \cite{shahzad2022permeability} performed numerical simulations to study the pressure losses of laminar flows through perforated plates. Here, the numerical simulations of three perforated plates are deployed to validate the proposed model. They have a same plate thickness ratio $\delta/D=0.25$ and different porosities $\varepsilon = 0.2$, $0.4$ and $0.5$. Figure \ref{fig:laminar_dp_with_Shahzad22} shows a comparison of the pressure drops predicted by both the present model (Eq. (\ref{parameters_improved})) and the Bae \& Kim model \cite{bae2016numerical} with the numerical simulations. Since the correction factor $f(\varepsilon, \delta/D)$ is less than $1$ for these perforated plates, the present model gives smaller predictions of pressure drops than the Bae \& Kim model \cite{bae2016numerical} does. This is more obvious for the plate with $\varepsilon = 0.2$ since the correction factor decreases with a decrease of plate porosity. It is also interesting to note that, the agreement between the present model and the Bae \& Kim model \cite{bae2016numerical} is better at $\varepsilon=0.4$ and $0.5$ than at $\varepsilon=0.2$, probably because the porosities of the perforated plates used to design the correction factor (Eq. (\ref{factor})) are between 0.4 and 0.63 and the perforated plate here with $\varepsilon=0.2$ is relatively far away from this porosity range. In other words, the present model probably has better prediction capabilities for plates with a moderate porosity that is not away far from the range between 0.4 and 0.63. Overall, compared with the numerical simulations of Shahzad et al. \cite{shahzad2022permeability}, both models give reasonable predictions of the pressure drops, demonstrating the capability of both models in predicting laminar flows through perforated plates. Indeed, since at low Reynolds numbers the Darcy contribution of pressure drop dominates over the Forchheimer contribution, the present model inherits the prediction capability from the Bae \& Kim model \cite{bae2016numerical}.

\subsection{Turbulent flow}\label{turbulent_flow}

\begin{figure}
\centerline{\includegraphics[angle=0,width=12.0cm]
  {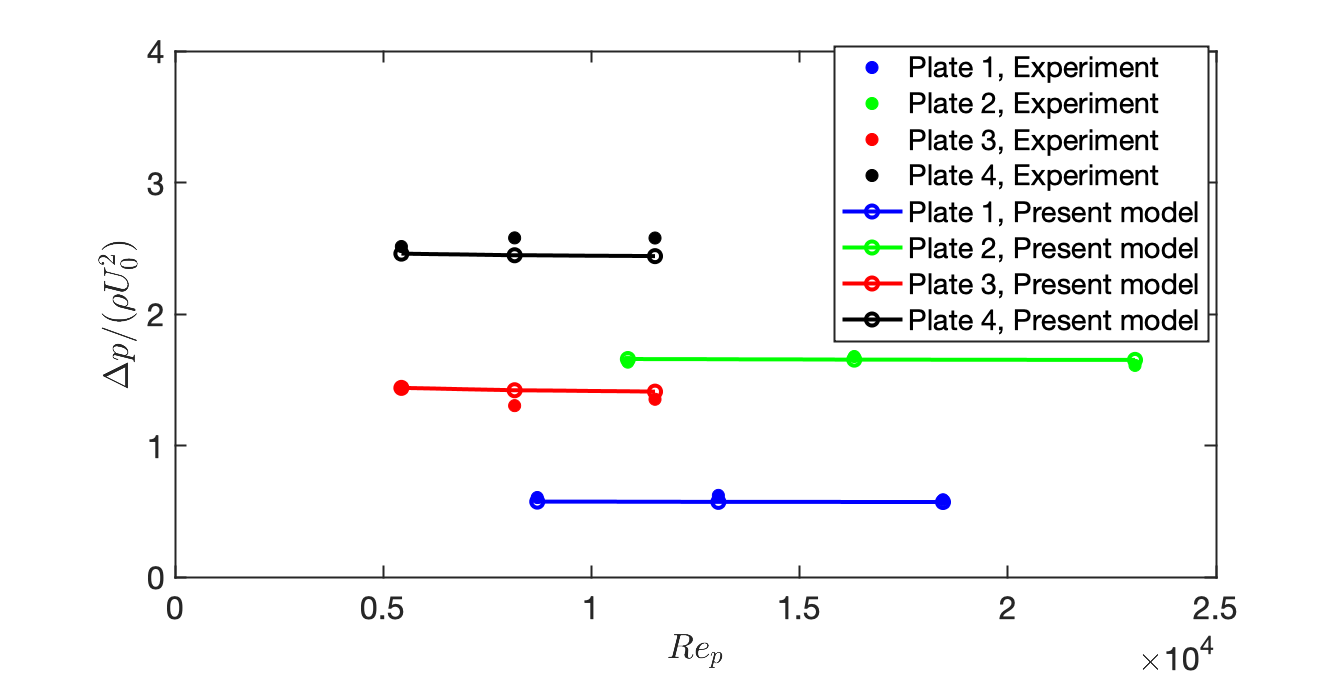}}
\caption{Comparison of the pressure drop predicted by the present model with the experiment of Méry \& Sebbane \cite{mery2023aerodynamic}.} 
\label{fig:dp_with_MS23}
\end{figure}

\begin{figure}
\centerline{\includegraphics[angle=0,width=12.0cm]
  {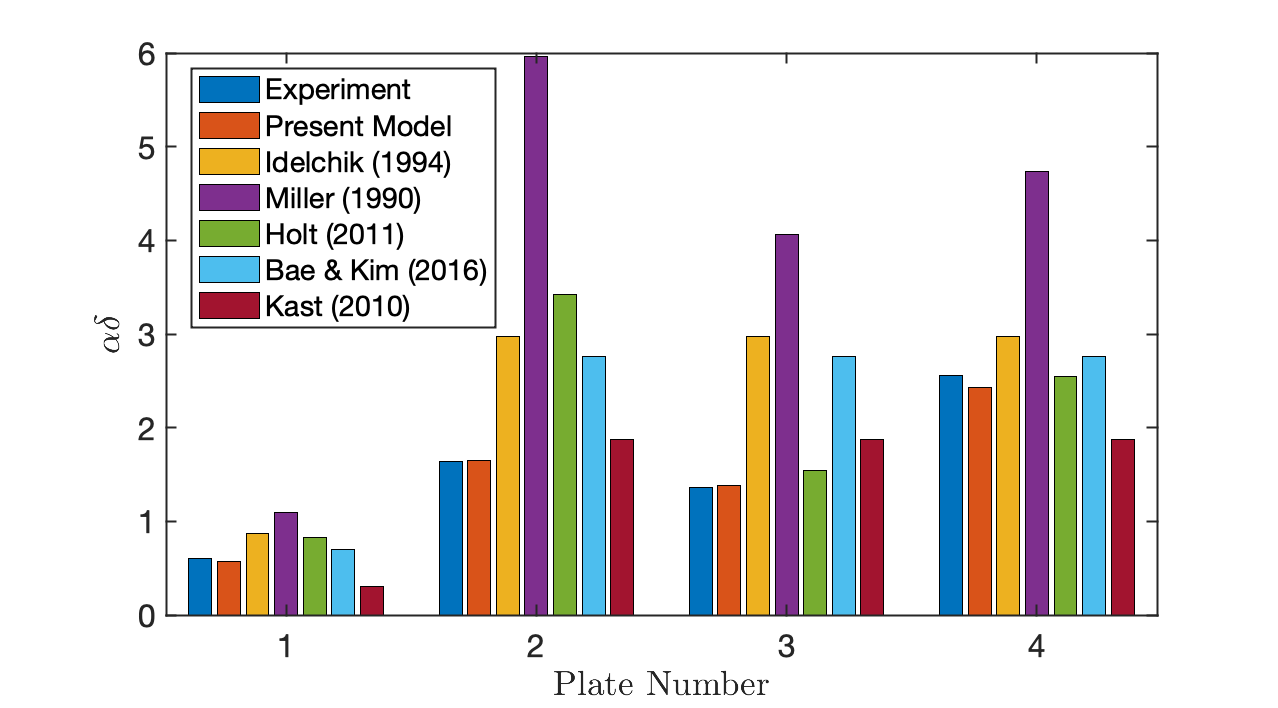}}
\caption{Comparison of the Forchheimer part of the pressure drop predicted by the present model with the measured pressure drops of Méry \& Sebbane \cite{mery2023aerodynamic} and the predictions by other previous models \cite{idelchik1994handbook, kast2010pressure, miller1990internal, holt2011cavitation}. Note that the pressure-drop values of the experiment are the average of the three values associated with the three different flow speeds for each plate.} 
\label{fig:dp_Forchheimer_MS23}
\end{figure}

Since the correction factor (Eq. (\ref{factor})) was proposed on account of the differences between the experimental data of Méry \& Sebbane \cite{mery2023aerodynamic} and the prediction of the Bae \& Kim model \cite{bae2016numerical}, we first validate the proposed model against the experimental data set of Méry \& Sebbane \cite{mery2023aerodynamic} in turbulent regimes. Figure \ref{fig:dp_with_MS23} shows a comparison of the pressure drop predicted by the present model with the experiment of Méry \& Sebbane \cite{mery2023aerodynamic}. With the correction factor in Eq. (\ref{factor}), the predictions of the new model agree much better with the experiment than those of the Bae \& Kim model \cite{bae2016numerical} (see Fig. \ref{fig:dp_by_Bae_Kim_16_with_MS23}). Figure \ref{fig:dp_Forchheimer_MS23} shows a comparison of the Forchheimer part of the predicted pressure drop with the experiment of Méry \& Sebbane \cite{mery2023aerodynamic} and the predictions by other previous models \cite{idelchik1994handbook, kast2010pressure, miller1990internal, holt2011cavitation}. The Forchheimer part basically represents the total pressure drop, since in turbulent regimes the Forchheimer contribution strongly dominates over the Darcy contribution, as shown in Fig. \ref{fig:dp_by_Bae_Kim_16_with_MS23}. According to Fig. \ref{fig:dp_Forchheimer_MS23}, it is evident that the present model gives better agreement with the experiment than other previous models do. However, it should be noted that the present comparisons with the experiment of Méry \& Sebbane \cite{mery2023aerodynamic} are not completely convincing to validate the new model since the proposed correction factor was based on the same set of experimental data. It is thus more convincing to demonstrate the prediction capability of the present model by validating the present model against another independent set of experimental data.

\renewcommand{\figurename}{Fig.}

\renewcommand{\figurename}{Fig.}

\begin{figure}
\centerline{\includegraphics[angle=0,width=12.0cm]
  {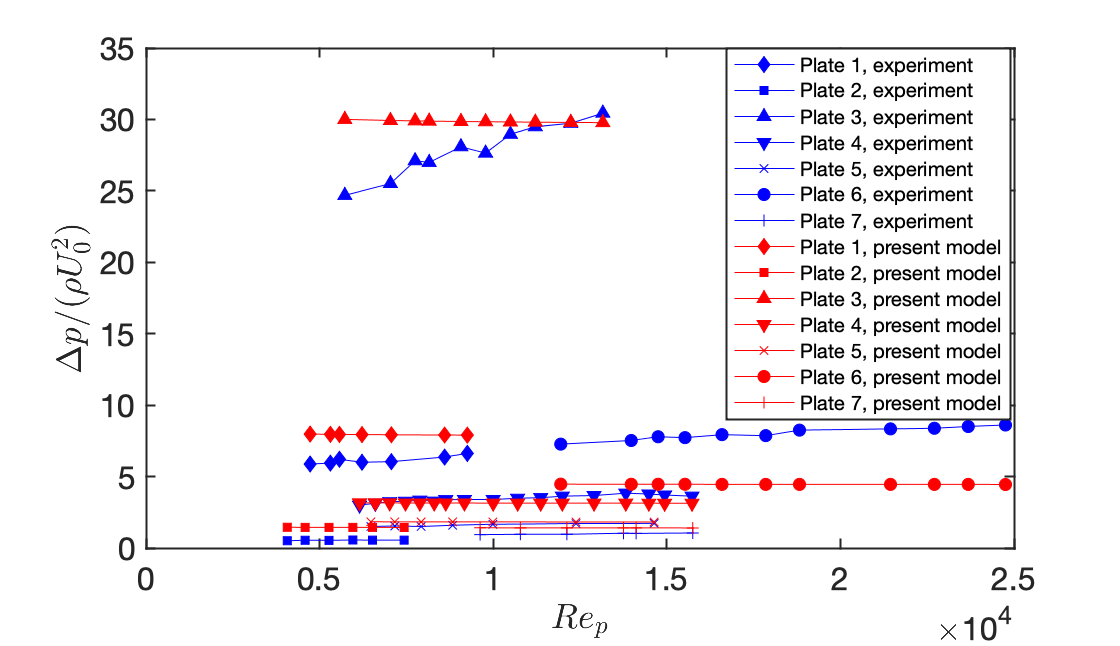}}
\caption{Comparison of the pressure drop predicted by the present model with the experiment of Yuvuzkurt \& Catchen \cite{yavuzkurt2003dependence}.} 
\label{fig:dp_with_YC03}
\end{figure}

Figure \ref{fig:dp_with_YC03} shows a comparison of the pressure drop predicted by the present model with the experiment of Yuvuzkurt \& Catchen \cite{yavuzkurt2003dependence}. Except for Plate 1, the measured pressure drops ($\Delta p/(\rho U_0^2)$) are nearly independent of the pore-level Reynolds number. However, for Plate 1, the dependence of pressure drop ($\Delta p/(\rho U_0^2)$) on the pore-level Reynolds number is approximately linear. Yuvuzkurt \& Catchen \cite{yavuzkurt2003dependence} attributed this to the effects of large-scale flow unsteadiness in the wake of the plate, which is characterized by a length scale that is much longer than the pore diameter or pore spacing. 

\begin{figure}
\centerline{\includegraphics[angle=0,width=12.0cm]
  {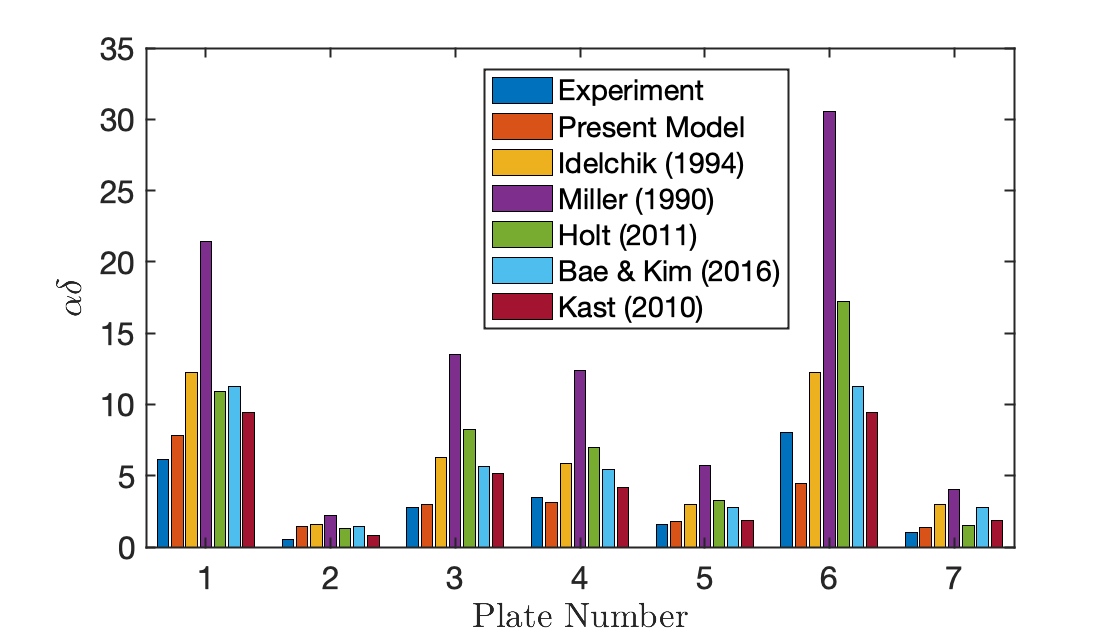}}
\caption{Comparison of the Forchheimer part of the pressure drop predicted by the present model with the measured pressure drops of Yuvuzkurt \& Catchen \cite{yavuzkurt2003dependence} and the predictions by other previous models \cite{idelchik1994handbook, kast2010pressure, miller1990internal, holt2011cavitation}. In this figure, for Plate 3, the results have been multiplied by 0.1 in order to have a better scale of the plot. Note also that the pressure-drop values of the experiment are the average of all values associated with different pore-level Reynolds numbers for each plate.} 
\label{fig:dp_Forchheimer_YC03}
\end{figure}

One the one hand, compared with the experiment, the present model gives reasonable predictions with relatively small discrepancies for Plates 2, 4, 5 and 7. Nevertheless, obvious discrepancies are observed for Plates 1, 3 and 6, again probably because the porosities of Plates 1, 3 and 6 are relatively far away from the porosity range between 0.4 and 0.63 (see Table \ref{tab:Yuvuzkurt_Catchen}). It is also observed that Plate 4 with a porosity of 0.3 gives a fairly good prediction. This suggests that, as long as the plate porosity is not far away from the porosity range between 0.4 and 0.63, the plate porosity does not need to fall exactly into this porosity range in order for the present model to produce good predictions. On the other hand, compared with other previous models \cite{idelchik1994handbook, kast2010pressure, miller1990internal, holt2011cavitation}, Fig. \ref{fig:dp_Forchheimer_YC03} suggests that the present model overall produces better agreement with the experiment of Yuvuzkurt \& Catchen \cite{yavuzkurt2003dependence} than other models do. This is significant, considering that the present model is in a much simpler form than the models of Idelchik \cite{idelchik1994handbook}, Miller \cite{miller1990internal}, Holt et al. \cite{holt2011cavitation} and Kast \cite{kast2010pressure} (see Appendix). 

\renewcommand{\figurename}{Fig.}

\renewcommand{\figurename}{Fig.}

\section{Application of the present model in numerical simulations}\label{application_model}

\subsection{Numerical modeling of perforated plates in two-dimensional channel flows}

For perforated plates consisting of very fine pores, direct meshing of the perforated plates is very challenging since it would result in a large amount of computational cells. In such cases, a numerical model to represent the effects of flow-through perforated plates is favoured. Similar to the modeling of other porous medium \cite{okolo2017numerical, zhu2020numerical, terracol2021numerical, li2023numerical}, the effect of perforated plates in flows can be modeled by adding a pressure-drop source term to the momentum equations. Within the perforated-plate region, a source term S is added to the right-hand side (RHS) of the momentum equations, yielding the RANS momentum equations as \cite{ccmUG}
\begin {equation} 
\frac{\partial (\rho \bar{u}_i)}{\partial t} + \frac{\partial (\rho \bar{u}_i\bar{u}_j)}{\partial x_j}  = - \frac{\partial \bar{p}}{\partial x_i} + \frac{\partial \bar{\sigma}_{ij}}{\partial x_j} + \frac{\partial \sigma_{ij, RANS}}{\partial x_j} + S_i,
\end {equation}
where $\bar{u}_i$ is the components of velocity (for two-dimensional flows, $i=$1, 2), $\bar{p}$ is the pressure, $\bar{\sigma}_{ij}$ is the viscous stress tensor and $\sigma_{ij, RANS}$ is the Reynolds-stress tensor.

If the two-dimensional air flow approaches the perforated plate at zero incident angle, the source term is expressed as
\begin{equation}\label{source_term}
     \begin{bmatrix}
           S_{1} \\
           S_{2} 
      \end{bmatrix} =   \begin{bmatrix}
                                      -\mu\bar u/K- \rho \alpha \left|\bar u\right|\bar u \\
                                      0
                                 \end{bmatrix},
\end{equation}
where $\bar{u}$ is the streamwise velocity. The parameters $K$ and $\alpha$ are given by the present model in Eq. (\ref{parameters_improved}).

\subsection{Computational meshes and simulation set-up}

\begin{figure}
\centerline{\includegraphics[angle=0,width=10.0cm]
  {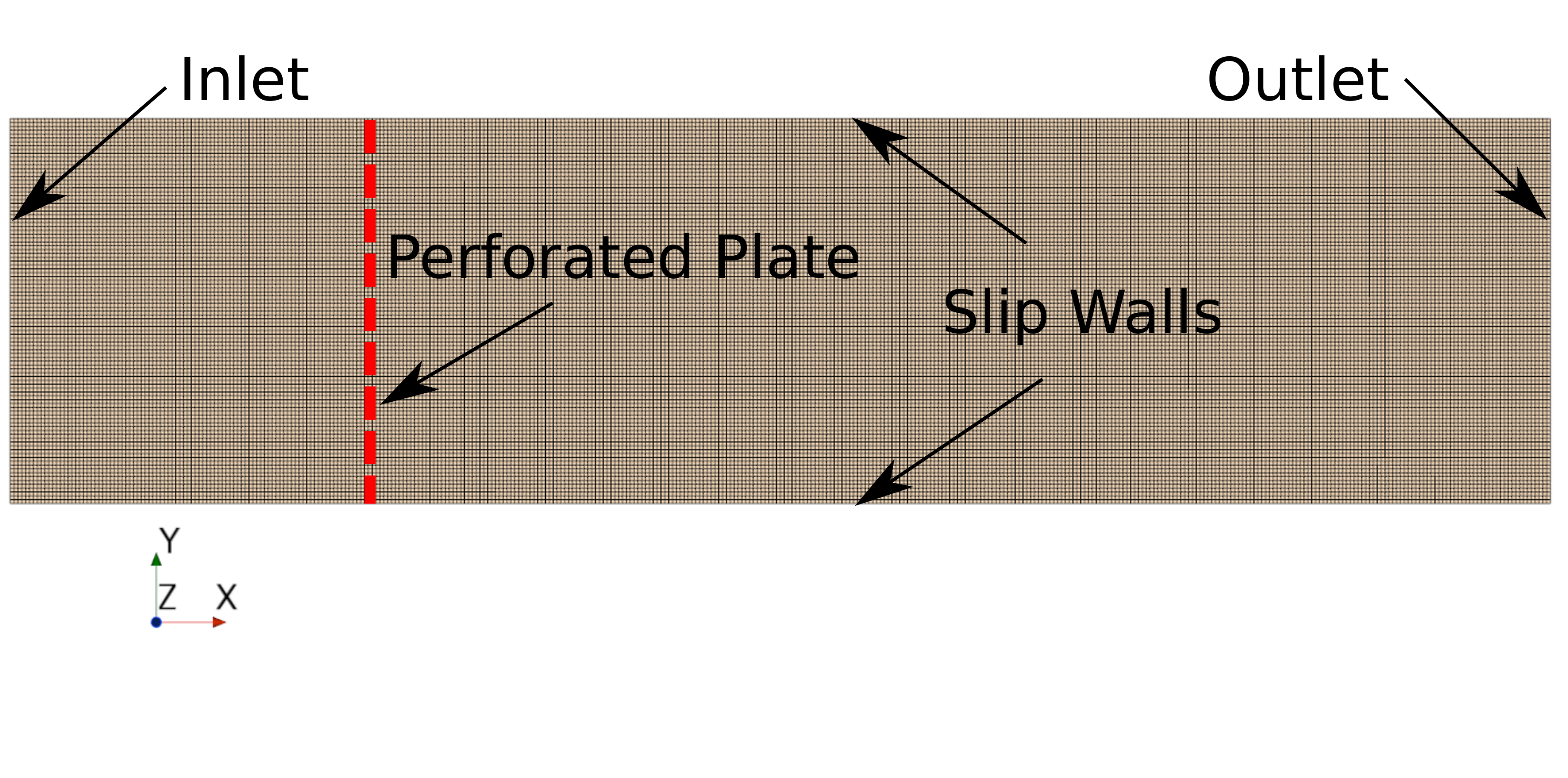}}
\caption{The two-dimensional coarse mesh with a cell size of $\Delta=0.5$ mm for the numerical simulations. The dashed line indicates the location where the source term is applied in the momentum equations.} 
\label{fig:Mesh_2D}
\end{figure}

The computational domain is a rectangle of $200 \times 50 \: \mathrm{mm^2}$. Three different computational meshes consisting of square cells of different size are deployed in the simulations. The coarse mesh has a cell size of $\Delta = 0.5$ mm, while the medium and fine meshes have a cell size of $\Delta = 0.25$ and $0.125$ mm, respectively. The cell numbers of the coarse, medium and fine meshes are 40k, 160k and 640k, respectively. Fig. \ref{fig:Mesh_2D} shows the coarse computational mesh. The perforated plate is not meshed, whereas its effect is modeled by adding a momentum source (see Eq. (\ref{source_term})) at the location of the dashed line ($x=0$). More specifically, for Plates 1, 2 and 4 with a thickness of $1$ mm, the momentum source is applied to the cells within $-0.5 \: \mathrm{mm} < x < 0.5 \: \mathrm{mm}$. For Plate 3 with a thickness of $2$ mm, the momentum source is applied to the cells within $-1 \: \mathrm{mm} < x < 1 \: \mathrm{mm}$. 

We perform steady RANS simulations with the SST k-$\omega$ model using the commercial Siemens STAR-CCM+ version 2021.1 \cite{ccm0}. The inlet velocity is $U_0 = 16.6$ m/s, leading to a Reynolds number of $Re=27370$ based on the inlet velocity and the channel half height. At the inlet boundary, the turbulence intensity is 6.2\% and the turbulent viscosity ratio is 10.0 which gives the dissipation rate $\omega$.

\renewcommand{\figurename}{Fig.}

\subsection{Pressure drop predicted by RANS simulations with the present model as a momentum source}

\begin{figure}
\centerline{\includegraphics[angle=0,width=12.0cm]
  {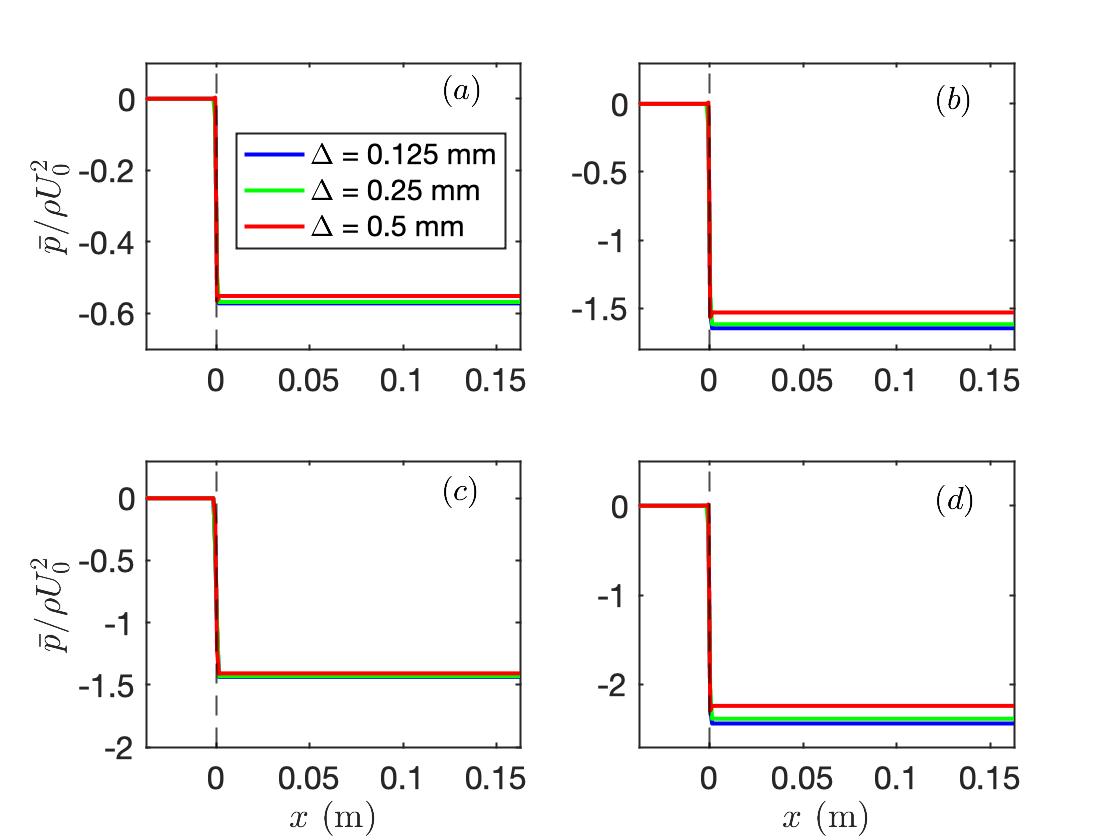}}
\caption{The RANS-predicted pressure drops using the coarse, medium and fine meshes. (\textit{a}) Plate 1; (\textit{b}) Plate 2; (\textit{c}) Plate 3; and (\textit{d}) Plate 4. The vertical black dashed line indicates the location ($x=0$) where the source term is applied in the momentum equations.} 
\label{fig:dp_RANS}
\end{figure}

\begin{figure}
\centerline{\includegraphics[angle=0,width=12.0cm]
  {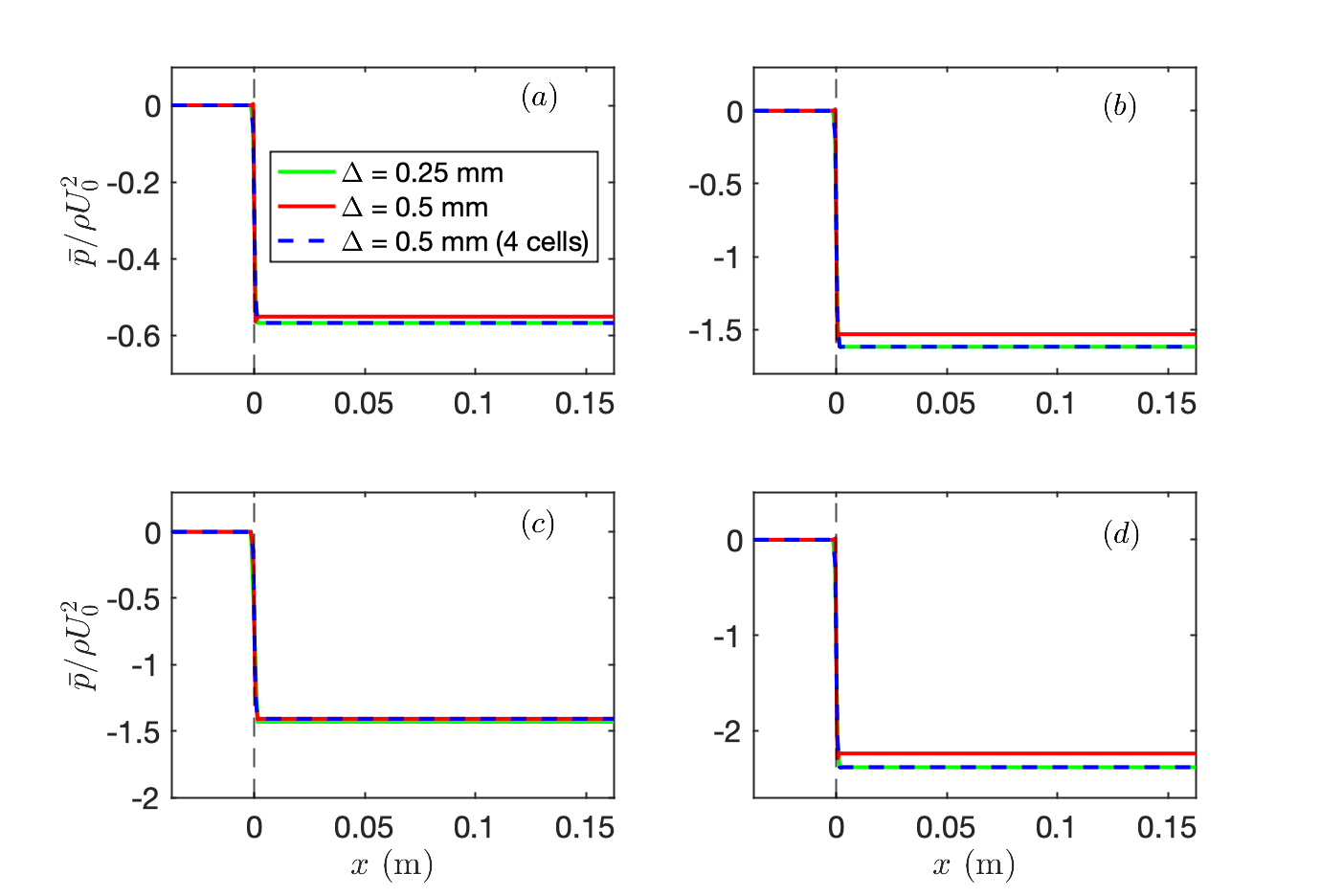}}
\caption{The RANS-predicted pressure drops using the coarse and medium meshes.  For the coarse mesh, simulations are also performed using 4 cells in the streamwise direction to represent the perforated plate. (\textit{a}) Plate 1; (\textit{b}) Plate 2; (\textit{c}) Plate 3; and (\textit{d}) Plate 4. The vertical black dashed line indicates the location ($x=0$) where the source term is applied in the momentum equations.} 
\label{fig:dp_RANS_oneWith4cells}
\end{figure}

Figure \ref{fig:dp_RANS} shows the pressure drops predicted by RANS simulations. The convergence of the solutions using the medium and fine meshes is well demonstrated. However, the coarse mesh ($\Delta = 0.5$ mm) seems to be too coarse for accurately predicting the pressure drops of Plates 1,  2 and 4. Meanwhile, it is also observed that the coarse mesh ($\Delta = 0.5$ mm) gives a fairly good prediction for Plate 3. To understand the reason, for Plates 1,  2 and 4, we perform additional simulations using the same coarse mesh but with the source term applied to the cells within $-1 \: \mathrm{mm} < x < 1 \: \mathrm{mm}$ instead of $-0.5 \: \mathrm{mm} < x < 0.5 \: \mathrm{mm}$, reducing the pressure drop in each cell by 50\% in order to keep the overall resistance the same. As such, all four plates are represented by 4 cells in the streamwise direction because the cell size is $\Delta=0.5$ mm. Figure \ref{fig:dp_RANS_oneWith4cells} shows the simulation results. It is clear that the solutions using 4 cells in the streamwise direction to represent the perforated plates converge to the solutions of the medium mesh ($\Delta=0.25$ mm). This suggests that, when modeling perforated plates (and probably also other porous plates), the modeling thickness of plate, whether the same as the physical thickness of plate or not, is less important. Rather, the number of cells used to represent the plate is more important \cite{li2023numerical}. It is suggested to have at least 4 cells in the streamwise direction to represent the flow through perforated plates.

The RANS-predicted pressure drops are also listed in Table \ref{tab:comparison} for a comparison with the measured pressure drops and the model-predicted pressure drops. The results pertaining to the simulations on the medium mesh using 8 cells in the streamwise direction to represent the perforated plates are also listed in the table. Overall, the RANS predictions agree very well with the present model predictions. Note that we also perform inviscid flow simulations. The results are the same as the RANS simulations. For brevity, the results associated with the inviscid flow simulations are not shown here.

\renewcommand{\figurename}{Fig.}

\renewcommand{\figurename}{Fig.}

\begin{table}[width=.45\linewidth,cols=4,pos=h]
\begin{threeparttable}
\caption{Comparison between the measured pressure drops of Méry \& Sebbane \cite{mery2023aerodynamic} and the pressure drops predicted by the present perforated-plate model and by numerical simulations using the medium mesh.}\label{tab:comparison}
\begin{tabular*}{\tblwidth}{@{} ccccc@{} }
\toprule
      Plate  & 1 & 2 & 3 & 4  \\[3pt]
\midrule
       Experiment   & 0.605 & 1.635 & 1.440 & 2.515 \\
       Present model   & 0.574 & 1.658 & 1.438 & 2.459 \\ 
       RANS Prediction   & 0.568 & 1.618 & 1.430 & 2.384 \\ 
       RANS Prediction\tnote{a}   & 0.572 & 1.647 & 1.430 & 2.436 \\        
\bottomrule
\end{tabular*}
    \begin{tablenotes}
      \item[a]{There are 8 cells in the streamwise direction for all 4 plates.}
    \end{tablenotes}
\end{threeparttable}
\end{table}

\section{Concluding remarks}\label{conclusion}

A fluid flow through a perforated plate is a common problem in a wide variety of practical applications in thermal, mechanical, chemical, civil, nuclear, ocean and aerospace engineering.  In this paper, we proposed a novel fluid flow model for the pressure loss through perforated plates in both laminar and turbulent flows. The design of this model is based on the recent experimental data of Méry \& Sebbane \cite{mery2023aerodynamic}, as well as the existing model of Bae \& Kim \cite{bae2016numerical} for laminar flows. 

The pressure losses caused by flow-through perforated plates are often characterized by the Darcy-Forchheimer equation. The Darcy term corresponds to the viscous pressure loss while the Forchheimer term represents the inertial pressure loss. It has been shown by Bae \& Kim \cite{bae2016numerical} that, at low Reynolds numbers ($Re_p<25$), the Darcy contribution to the normalized pressure drop dominates over the Forchheimer contribution. In turbulent regimes, however, it is found that the Forchheimer contribution strongly dominates. Moreover, the existing model of Bae \& Kim \cite{bae2016numerical} was only applicable to laminar flows. For these reasons, we proposed a correction to the Forchheimer coefficient of the Bae \& Kim model \cite{bae2016numerical}  to enable the modeling of perforated plates in both laminar and turbulent flows. The new model is then validated against existing numerical simulations in the laminar regime and experiments in the turbulent regime. Overall, the predictions given by the new model agree well with the numerical simulations and experiments, and are superior to other models in the literature \cite{idelchik1994handbook, kast2010pressure, miller1990internal, holt2011cavitation}. This is significant, considering that the present model is much simpler than the models of Idelchik \cite{idelchik1994handbook}, Miller \cite{miller1990internal}, Holt et al. \cite{holt2011cavitation} and Kast \cite{kast2010pressure} (see Appendix). 

To demonstrated the application of the new model in numerical simulations, two-dimensional channel flows are simulated using RANS with the new model as a pressure-drop source term added to the momentum equations. Overall, the RANS predictions agree very well with the present model predictions. In the numerical simulations, it is suggested to have at least 4 cells in the streamwise direction to represent the flow-through perforated plates.

Although the present model is applicable for a wide range of Reynolds numbers ranging from laminar to turbulent regimes, the limitation of the model is that the new model is only applicable for a certain range of plate porosities ($\varepsilon$) and plate thickness ratios ($\delta/D$). This is limited by the experimental data used for the design of the present model. On the one hand, the model is applicable to a plate thickness ratio of $0.2<\delta/D<1$ because the correction factor $f$ would go negative as $\delta/D>1.2$. On the other hand, the present model probably has good prediction capabilities for plates with a moderate porosity that is not away far from the range between 0.4 and 0.63. The authors suggest an applicable porosity range between 0.3 and 0.7. Fortunately, these ranges of plate porosities and thicknesses have enabled the use of perforated plates in numerous engineering problems, e.g. heat transfer in heat exchangers, flame control in combustion chambers, aerodynamic noise abatement, etc.

\appendix
\section*{Appendix: Empirical formulae for the pressure loss through perforated plates}

Besides the aforementioned model of Bae \& Kim \cite{bae2016numerical} which was designed using numerical simulations of laminar flows, here we append some other popular models (formulae) for estimating the Forchheimer coefficient $\alpha$. These formulae were proposed in experimental studies at high Reynolds numbers where the Forchheimer term strongly dominates over the Darcy term. Thus, the Forchheimer coefficient $\alpha$ can be used to calculate the Forchheimer contribution which is approximately equal to the pressure loss.

For perforated plates of finite thickness at high Reynolds numbers, Idelchik \cite{idelchik1994handbook} proposed a formula of the form
\begin {equation}\label{Idelchik1994}
\alpha = \frac{1}{2\varepsilon^2\delta}\left(0.5+0.24\sqrt{1-\varepsilon}(1-\varepsilon)+(1-\varepsilon)^2\right).
\end {equation}

Similarly, Kast \cite{kast2010pressure} suggested the following formula
\begin {equation}\label{Kast2010}
\alpha = \frac{1}{2\varepsilon^2\delta}\left((\frac{1}{C}-1)^2+(1-\varepsilon)^2\right),
\end {equation}
where $C$ is a coefficient depending upon the plate porosity and can be expressed as
\begin {equation}
C = 0.6 + 0.4\varepsilon^2.
\end {equation}

Miller \cite{miller1990internal} proposed the following formula 
\begin {equation}
 \alpha = C_0\frac{(1-C_c\varepsilon)^2}{C_c^2\varepsilon^2\delta},
\end {equation}
where $C_0$ is a coefficient that depends on $\delta/D$, and $C_c$ is the jet contraction coefficient that depends on $\varepsilon$. According to Fratino \cite{fratino2000hydraulic}, $C_0$ can be calculated by the following empirical expression, valid for $0.1<\delta/D<3$,
\begin {equation}
C_0 = 0.5 + \frac{0.178}{4(\frac{\delta}{D})^2+0.355},
\end {equation}
while $C_c$ is given by
\begin {equation}
C_c = 0.596 + 0.0031\exp(\frac{\sqrt{\varepsilon}}{0.206}).
\end {equation}

Holt et al. \cite{holt2011cavitation} proposed a piecewise function for the Forchheimer coefficient, as follows
\begin {equation} 
\label{convective_term_BCD}
\alpha = 
    \begin{cases}
        \frac{1}{2\delta}\left(2.9-3.79\frac{\delta}{D}\varepsilon^{0.2}+1.79\frac{\delta}{D}^2\varepsilon^{0.4}\right)K_{LA}  & \frac{\delta}{D}\varepsilon^{0.2}<0.9 \\
        \frac{1}{2\delta}\left(0.876+0.069\frac{\delta}{D}\varepsilon^{0.2}\right)K_{LA}  & \hfill \break \frac{\delta}{D}\varepsilon^{0.2}>0.9
    \end{cases}
\end {equation}
where the jet contraction coefficient $C_c$ is set to be 0.72 \cite{malavasi2012pressure}. $K_{LA}$ is the pressure loss coefficient of a single-hole orifice as estimated by a theoretical model for reattached flows, as follows
\begin {equation}
K_{LA} = 1 - \frac{2}{\varepsilon} + \frac{2}{\varepsilon^2}\left(1-\frac{1}{C_c}+\frac{1}{2C_c^2}\right).
\end {equation}

\printcredits

\section*{Declaration of Competing Interest}
The authors declare that they have no known competing financial interests or personal relationships that could have appeared to influence the work reported in this paper.

\section*{Data Availability}
The data that support the findings of this study are available from the corresponding author upon reasonable request.

\section*{Funding Source}
This work is supported by the project Innovative Design of Installed Airframe Components for Aircraft Noise Reduction ("INVENTOR", European Union’s Horizon 2020 Research and Innovation Programme, under Grant Agreement N$^\circ$ 8605383). 

\section*{Acknowledgments}
We would like to acknowledge Dassault Aviation (Dr. Vincent Fleury) for defining the parameters (in Table 1) of the perforated plates used in the B2A wind-tunnel experiment. The computations were enabled by the computer resources at the Chalmers Centre for Computational Science and Engineering (C3SE) provided by the Swedish National Infrastructure for Computing (SNIC). 
\bibliographystyle{elsarticle-num}

\bibliography{cas-refs}


\end{document}